\documentclass{article}
\usepackage{cite}

\usepackage[english]{babel}

\usepackage[table]{xcolor}
\usepackage{tabularx,booktabs}
\usepackage{colortbl}
\usepackage{amsmath}
\usepackage{graphicx}
\usepackage{subcaption}
\usepackage[colorlinks=true, allcolors=blue]{hyperref}
\usepackage{longtable}
\usepackage{pgfplots}
\usepackage{pgfplotstable}
\usepackage{tikz}
\usepackage{authblk}
\usepackage[a4paper, margin=1.2in]{geometry}
\pgfplotsset{compat=1.18}
\usepackage[table]{xcolor}
\usepackage{colortbl}
\usepackage{amsmath}
\usepackage{graphicx}
\usepackage{subcaption}
\usepackage[colorlinks=true, allcolors=blue]{hyperref}
\usepackage{longtable}
\usepackage{tikz}
\usepackage{amsmath}
\usepackage{graphicx} 
\usepackage[inline]{enumitem}
\usetikzlibrary{shadows.blur, shapes.geometric, positioning, chains, shapes.symbols}
\usetikzlibrary{arrows.meta, positioning}

\definecolor{LBBlue}{HTML}{66b1ff}   
\definecolor{LBRed}{HTML}{ff7f7f}    
\definecolor{LBGreen}{HTML}{7fd37f}  
\definecolor{color1}{HTML}{fafafa}
\definecolor{color2}{HTML}{e4e5f1}
\definecolor{color3}{HTML}{d2d3db}
\definecolor{color4}{HTML}{9394a5}
\definecolor{myorange}{HTML}{ffa729}
\definecolor{mygray}{HTML}{969696}
\definecolor{mygreen}{HTML}{1ed200}

\newcommand{\circlecolor}[2][1ex]{%
  \tikz\draw[fill=#2, draw=none] (0,0) circle [radius=#1];%
}
\begin{document}

\title{Inside LockBit: Technical, Behavioral, and Financial Anatomy of a Ransomware Empire}

\providecommand{\keywords}[1]{\textbf{\textit{Keywords:}} #1}

\author[1,2]{Felipe Castaño\thanks{\url{fcastano@vicomtech.org}}}
\author[2,3]{Constantinos Patsakis\thanks{\url{kpatsak@unipi.gr}}}
\author[1]{Francesco Zola\thanks{\url{fzola@vicomtech.org}}}
\author[5,3]{Fran Casino\thanks{\url{franciscojose.casino@urv.cat}}}

\affil[1]{Digital Security Department, Vicomtech (BRTA), Donostia/San Sebastian, Spain}
\affil[2]{Dept. of Electrical Engineering, Systems and Automation, Universidad de León, León, Spain\thanks{\url{fcastano@vicomtech.org}}}
\affil[3]{Department of Informatics, University of Piraeus, Piraeus, Greece}
\affil[4]{Athena Research Centre, Greece}
\affil[5]{Dept. of Computer Engineering and Mathematics, Universitat Rovira i Virgili, Catalonia, Spain}

\date{}
\maketitle

\begin{abstract}
LockBit has evolved from an obscure Ransomware-as-a-Service newcomer in 2019 to the most prolific ransomware franchise of 2024. Leveraging a recently leaked MySQL dump of the gang's management panel, this study offers an end-to-end reconstruction of LockBit's technical, behavioral, and financial apparatus. We recall the family's version timeline and map its tactics, techniques, and procedures to MITRE ATT\&CK, highlighting the incremental hardening that distinguishes LockBit 3.0 from its predecessors. We then analyze 51 negotiation chat logs using natural-language embeddings and clustering to infer a canonical interaction playbook, revealing recurrent rhetorical stages that underpin the double-extortion strategy. Finally, we trace 19 Bitcoin addresses related to ransom payment chains, revealing two distinct patterns based on different laundering phases. In both cases, a small portion of the ransom is immediately split into long-lived addresses (presumably retained by the group as profit and to finance further operations) while the remainder is ultimately aggregated into two high-volume addresses before likely being sent to the affiliate. These two collector addresses appear to belong to distinct exchanges, each processing over 200k BTC. The combined evidence portrays LockBit as a tightly integrated criminal service whose resilience rests on rapid code iteration, script-driven social engineering, and industrial-scale cash-out pipelines.   
\end{abstract}


\section{Introduction}

High-Risk Criminal Networks always exploit emerging communication platforms and various tactics to fund their illegal activities, including the Crime-as-a-Service (CaaS) strategy, which leverages cooperation and specialization among cybercriminals to increase the complexity and impact of their operation \cite{cepol1}. In this scenario, ransomware represents one of the most demanded and deployed threats, primarily due to its ability to affect large populations and generate huge economic returns \cite{europol2024iocta}. Furthermore, the as-a-service paradigm enables various ransomware strains to adopt similar mechanisms for malware creation, distribution, ransom collection, and money laundering \cite{zola2024unveiling}, favoring the creation of the Ransomware-as-a-Service (RaaS) \cite{alwashali2021survey, meland2020ransomware}. This allows the ransomware provider to speed up the mass production of new strains and, at the same time, enables service applicants to rely on an innovative system based on a solid and proven \emph{modus operandi}. 

This coordinated methodology, in which affiliates operate under guidance provided by core developers or manuals, increases the professionalization of cybercrime. Thus, it contributes to the repeatability and effectiveness of attacks on different targets. In this context, it is reasonable to hypothesize that ransomware groups (and affiliates) receive training or follow operational playbooks that guide them in how to contact, respond to, and negotiate with victims. These interactions appear to follow a predetermined process, allowing attackers to systematically escalate threats, increase psychological pressure, and push victims toward payment \cite{brewer2016ransomware}. This structured approach not only maximizes the impact of their attacks but also facilitates the concealment of their tracks.

At the same time, once the funds are collected, it is reasonable to hypothesize that groups apply similar mechanisms for money laundering. Specifically, as demonstrated in several previous studies, the principal medium for ransom collection and laundering is cryptocurrencies \cite{conti2018economic,zimba2019economic}. In fact, these blockchain currencies have introduced novel channels for facilitating payments and laundering illicit proceeds, mainly due to the decentralized nature of these systems and unregulated markets \cite{eusocta1}. Among these digital assets, Bitcoin stands out as the most widely recognized and accessible cryptocurrency, not only due to its high market capitalization but also because it is relatively easy to acquire, even for individuals with limited technical expertise. This ease of access has contributed to its dominance in cyber-criminal payments\cite{chainalysis2025crypto}. 

Among the different ransomware strains, one of the most harmful and widely employed, with a notable number of victims claimed on its data leak site, is the LockBit family. This ransomware operates under the RaaS model \cite{meland2020ransomware,crowdstrike}, which can be considered a part of the Malware-as-a-Service model \cite{patsakis2024malware}. The group has been active since around 2019 and has evolved to one of the most widespread and disruptive ransomware threats. In this RaaS scheme, the core LockBit developers lease their ransomware to affiliates, other cybercriminals who carry out the intrusions, in exchange for a share of the profits. In particular, LockBit was reported to be the most widely deployed ransomware worldwide in 2022 \cite{cisa}, with victims across many sectors (finance, government, healthcare, etc.). Despite some variation in tactics by different affiliates, LockBit attacks typically follow a similar multi-stage pattern involving initial access, lateral movement, data exfiltration, file encryption, and a double extortion strategy. The double extortion strategy involves stealing sensitive data before encrypting the victim's systems, thereby increasing the pressure on the victims to pay. This means that even if victims can restore their files from backups, the attackers can threaten to publish the stolen data unless a ransom is paid \cite{double}.

LockBit was the subject of multiple international enforcement efforts throughout 2023 and 2024, culminating in Operation Cronos\cite{europol_lock2} - a coordinated action led by Europol, Eurojust, and agencies from 12 countries. The operation resulted in the seizure of LockBit's infrastructure and access to internal systems, tackling the criminal operations of the ransomware group. Furthermore, several national governments imposed sanctions on ransomware administrators and affiliates, undermining the group's operational and financial capabilities \cite{europol_lock1}. 
Nonetheless, LockBit keeps reviving from its ashes. In fact, according to the \emph{Annual Cyber Threat Monitor Report 2024} \cite{ncc2024} released by the NCC group and the \textit{Kaspersky Security Network data} \cite{secure2025}, despite law enforcement efforts, LockBit returned with a vengeance, relaunched its operations, and remained active throughout 2024.
This trend was confirmed in the recent \emph{Chainalysis 2025 Crypto Crime Report} \cite{chainalysis2025crypto}, which highlighted that, although the group showed a sharp revenue decrease (almost 80\%) due to the law enforcement operation, it was not completely shut down.

In early May 2025, the LockBit group was hacked by someone claiming to originate from Prague, and a MySQL dump of their admin panel was publicly shared\footnote{\url{https://github.com/D4RK-R4BB1T/Criminal-Leaks/tree/main}}. The leaked SQL database dump spans from December 18, 2024, to April 29, 2025, and contains detailed information on LockBit affiliates, victim organizations, chat transcripts, cryptocurrency wallets, and ransomware build configurations.

In this article, we leverage the leak of LockBit's management panel to shed light on ransomware operations and deliver a unified view of the group's \textit{modus operandi}. Specifically, we aim to reconstruct it by presenting the ransomware timeline and technical evolution, analyzing behavioral patterns, examining the conversations between the group and victims, and finally, LockBit's financial operations. Section \ref{sec:operation} recalls the family's release timeline, tracing its path from the 2019 ``ABCD'' prototype to the cross-platform LockBit 3.0, and compares the ATT\&CK technique sets of LockBit 2.0 and 3.0, highlighting how recent builds deepen defense-evasion and execution coverage. Then, Section \ref{sec:leak} provides an overview of the leaked data and its users' activity. Section \ref{sec:attackerbehavior} analyzes 51 leaked negotiation chats, highlighting a stable affiliate playbook applied to the victims, also presenting some interesting incidental findings.
Section~\ref{sec:crypto} links these conversations to cryptocurrency data and analyses on-chain activities, shedding light on LockBit's possible \textit{modus operandi} in ransom collection and money laundering operations. Section \ref{sec:conclusions} discusses the practical lessons that emerge from the analysis and provides future research lines.

\section{Operation and evolution of LockBit}
\label{sec:operation}
LockBit's attack chain follows a disciplined, stage-driven routine \cite{akinyemi2023analysis,understanding_lockbit}. Operators first obtain access by exploiting exposed RDP/VPN endpoints or replaying credentials obtained through social engineering campaigns, third-party breaches, and infostealers \cite{10.1145/3600160.3605047}. Next, scripted \texttt{PowerShell}/BAT loaders drop post-exploitation frameworks (e.g., Cobalt Strike, PowerShell Empire) and establish persistence via autostart registry keys. The intruders escalate privileges using token impersonation and credential-dumping tools, such as \textit{Mimikatz} \cite{el2020detecting}, and then push customized Group Policy Objects (GPO) that disable Windows Defender, delete shadow copies and logs, and impair host defences. Comprehensive reconnaissance enumerates files, shares, and domain trusts, after which the attack propagates laterally over the SMB using embedded credential lists or PsExec/GPO. Before launching the ransomware, the \textit{StealBit}/MEGA module \cite{stealbit} exfiltrates selected data to cloud storage for double-extortion leverage. Finally, a multithreaded ChaCha20–AES locker \cite{najm2018comparing} encrypts accessible assets, drops a note in a file of the form \texttt{\textless ID\textgreater.README.txt}, and replaces the desktop wallpaper, making recovery impossible without the attacker's decryption key. Table \ref{tab:lockbit-tactics} maps the previous steps with MITRE ATT\&CK\ tactics \cite{lockbit30}, which has analyzed LockBit 3.0 version.

\begin{table}[ht]
    \centering
    \caption{LockBit 3.0 operational steps mapped to key MITRE ATT\&CK\ tactics.}
    \label{tab:lockbit-tactics}
    \rowcolors{2}{}{gray!10}
    \begin{tabular}{p{5cm}p{9.5cm}}
        \toprule
        \textbf{Attack phase} & \textbf{Representative LockBit 3.0 actions \& tools} \\ \midrule
        Execution \& persistence & User-executed PowerShell/BAT stagers (T1204); autostart registry keys / scheduled tasks (T1547); Cobalt Strike \& PowerShell Empire beacons. \\
        Privilege escalation \& defence evasion & Token manipulation (T1134); credential dumping with \textit{Mimikatz}; custom GPOs to disable AV and delete shadow copies (T1562). \\
        Discovery \& lateral movement & File/share and domain enumeration (T1083, T1135); SMB self-propagation and PsExec/GPO lateral tool transfer (T1570). \\
        Exfiltration & \textit{StealBit}, FreeFileSync, or MEGA to exfiltrate archives over HTTPS (T1567). \\
        Impact & System-wide ChaCha20–AES encryption (T1486); service stop/defacement (T1489, T1491); ransom note and branded wallpaper deployment. \\ \bottomrule
    \end{tabular}
\end{table}

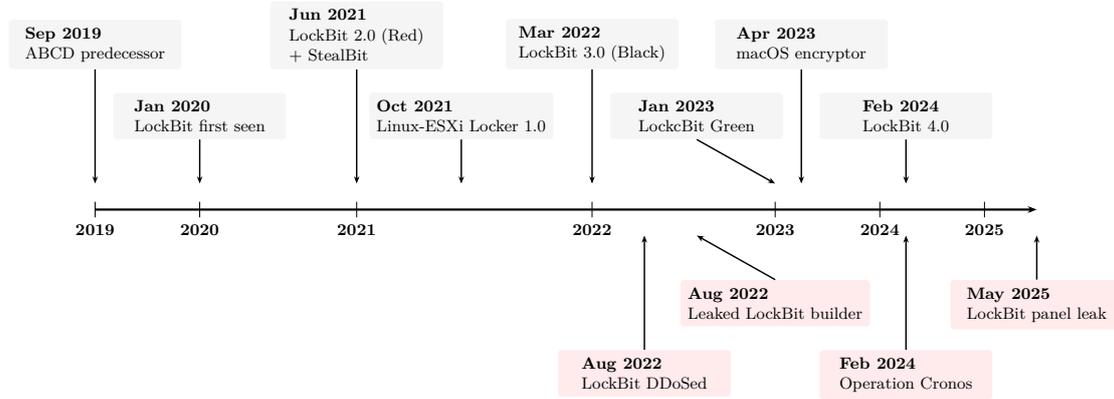
\begin{figure*}[!htbp]
\centering
\resizebox{\textwidth}{!}{ 
   \begin{tikzpicture}[
    timeline/.style   = {very thick, -{Stealth[length=5pt]}},
    event/.style      = {rounded corners=2pt, fill=gray!8,
                         font=\small, align=left,
                         minimum width=3.3cm, inner sep=4pt},
    eventAbove/.style = {event, anchor=south, yshift=2.7cm},
    eventBelow/.style = {event, anchor=south, yshift=1.35cm},
    year/.style       = {font=\bfseries\small, anchor=north},
    eventCritical/.style = {event, anchor=north, yshift=-1.35cm, fill=red!8},
    year/.style       = {font=\bfseries\small, anchor=north},
    eventCriticalB/.style = {event, anchor=north, yshift=-2.7cm, fill=red!8},
    year/.style       = {font=\bfseries\small, anchor=north},
    arrowToTimeline/.style = {-{Stealth[length=4pt]}, thick}
]

\draw[timeline] (0,0) -- (18,0);


\node[eventAbove] (e1) at (0,0)     {\textbf{Sep 2019}\\ABCD predecessor};
\draw[arrowToTimeline] (e1.south) -- (0,0.5);

\node[eventBelow] (e2) at (2,0)     {\textbf{Jan 2020}\\LockBit first seen};
\draw[arrowToTimeline] (e2.south) -- (2,0.5);

\node[eventAbove] (e3) at (5,0)     {\textbf{Jun 2021}\\LockBit 2.0 (Red)\\+ StealBit};
\draw[arrowToTimeline] (e3.south) -- (5,0.5);

\node[eventBelow] (e4) at (7,0)     {\textbf{Oct 2021}\\Linux-ESXi Locker 1.0};
\draw[arrowToTimeline] (e4.south) -- (7,0.5);

\node[eventAbove] (e5) at (9.5,0)   {\textbf{Mar 2022}\\LockBit 3.0 (Black)};
\draw[arrowToTimeline] (e5.south) -- (9.5,0.5);

\node[eventBelow] (e7) at (11.5,0)    {\textbf{Jan 2023}\\LockcBit Green};
\draw[arrowToTimeline] (e7.south) -- (13,0.5);

\node[eventAbove] (e8) at (13.5,0)    {\textbf{Apr 2023}\\macOS encryptor};
\draw[arrowToTimeline] (e8.south) -- (13.5,0.5);

\node[eventBelow] (e12) at (15.5,0)    {\textbf{Feb 2024}\\LockBit 4.0};
\draw[arrowToTimeline] (e12.south) -- (15.5,0.5);

\node[eventCriticalB] (e12) at (10.5,0)    {\textbf{Aug 2022}\\LockBit DDoSed};
\draw[arrowToTimeline] (e12.north) -- (10.5,-0.5);

\node[eventCritical] (e13) at (13,0)    {\textbf{Aug 2022}\\Leaked LockBit builder};
\draw[arrowToTimeline] (e13.north) -- (11.5,-0.5);

\node[eventCriticalB] (e10) at (15.5,0)    {\textbf{Feb 2024}\\Operation Cronos};
\draw[arrowToTimeline] (e10.north) -- (15.5,-0.5);

\node[eventCritical] (e9) at (18,0)    {\textbf{May 2025}\\LockBit panel leak};
\draw[arrowToTimeline] (e9.north) -- (18,-0.5);

\foreach \x/\lbl in {0/2019, 2/2020, 5/2021, 9.5/2022, 13/2023, 15/2024, 17/2025}
    \draw (\x,0.15) -- (\x,-0.15) node[year] {\lbl};

\end{tikzpicture}
   }
    \caption{Timeline of the main milestones of LockBit.}
    \label{fig:timeline}
\end{figure*}

In terms of evolution, LockBit has been continuously upgraded and revised by checking the tactics, techniques, and procedures (TTPs) used to deploy and execute ransomware. A timeline summary of the main versions and milestones is shown in Figure \ref{fig:timeline}. In this regard, when analyzing LockBit 2.0 and 3.0, we can understand the refinement of the latter in terms of defense evasion and exfiltration tactics and more robust encryption methods than its predecessor. Table \ref{tab:lb-coverage} provides a detailed description of the evolution of the MITRE ATT\&CK\ tactics of LockBit, according to MITRE  \cite{lockbit20,lockbit30}. As it can be observed, while more than 25 enterprise techniques are shared by LockBit 2.0 and 3.0, the latter exhibits more complex attack techniques. For instance, LockBit 3.0 can bypass User Account Control (UAC) to execute code with elevated privileges through an elevated Component Object Model (COM) interface. Moreover, LockBit 3.0 extends the encryption capabilities of LockBit 2.0, as it can encrypt targeted data using the AES-256, ChaCha20, or RSA-2048 algorithms. In addition, LockBit 3.0 can also use PowerShell to apply Group Policy changes, and it can install system services for persistence, as seen in Table \ref{tab:lockbit-tactics}. 

\begin{table}[!htbp]
\centering
\caption{Comparison of MITRE ATT\&CK enterprise techniques used by LockBit 2.0 and 3.0 according to MITRE  \cite{lockbit20,lockbit30}. \textbf{Color legend:} \colorbox{LBBlue}{\strut blue} — technique appears in \emph{both} LockBit 2.0 and 3.0; \colorbox{LBRed}{\strut red} — only in LockBit 2.0; \colorbox{LBGreen}{\strut green} — only in LockBit 3.0.}
\label{tab:lb-coverage}
\renewcommand{\arraystretch}{1.15}
\scriptsize
\begin{subtable}[t]{.45\textwidth}
\begin{tabular}{ll >{\raggedright\arraybackslash}p{6cm}}
\toprule
\textbf{ID} & \textbf{Name} \\ \midrule
 \cellcolor{LBBlue}T1021 & Remote Services \\
 \cellcolor{LBBlue}T1021.002 & SMB/Windows Admin Shares \\
 \cellcolor{LBGreen}T1027 & Obfuscated/Stored Files and Information \\
 \cellcolor{LBGreen}T1027.002 & Software Packing \\
 \cellcolor{LBGreen}T1027.013 & Encrypted/Encoded File \\
 \cellcolor{LBRed}T1047 & Windows Management Instrumentation \\
 \cellcolor{LBRed}T1053 & Scheduled Task/Job \\
 \cellcolor{LBRed}T1053.005 & Scheduled Task \\
 \cellcolor{LBBlue}T1057 & Process Discovery \\
 \cellcolor{LBBlue}T1059 & Command \& Scripting Interpreter \\
 \cellcolor{LBBlue}T1059.001 & PowerShell \\
 \cellcolor{LBRed}T1059.003 & Windows Command SheCommandll \\
 \cellcolor{LBBlue}T1070 & Indicator Removal on Host \\
 \cellcolor{LBBlue}T1070.001 & Clear Windows Event Logs \\
 \cellcolor{LBBlue}T1070.004 & File Deletion \\
 \cellcolor{LBGreen}T1071 & Application-Layer Protocol \\
 \cellcolor{LBGreen}T1071.001 & Web Protocols \\
 \cellcolor{LBGreen}T1078 & Valid Accounts \\
 \cellcolor{LBGreen}T1078.003 & Local Account \\
 \cellcolor{LBBlue}T1082 & System Information Discovery \\
 \cellcolor{LBBlue}T1083 & File \& Directory Discovery \\
 \cellcolor{LBGreen}T1106 & Native API \\
 \cellcolor{LBBlue}T1112 & Modify Registry \\
 \cellcolor{LBBlue}T1120 & Peripheral Device Discovery \\
 \cellcolor{LBGreen}T1132 & Data Encoding \\
 \cellcolor{LBGreen}T1132.001 & Standard Encoding \\
 \cellcolor{LBBlue}T1135 & Network Share Discovery \\
 \cellcolor{LBRed}T1136 & Create Account \\
 \cellcolor{LBBlue}T1140 & Deobfuscate/Decode Files/Info \\
\bottomrule
\end{tabular}
\end{subtable}\hfill
\begin{subtable}[t]{.45\textwidth}
\begin{tabular}{ll >{\raggedright\arraybackslash}p{6cm}}
\toprule
\textbf{ID} & \textbf{Name} \\ \midrule
 \cellcolor{LBGreen}T1218 & Signed Binary Proxy Execution \\
 \cellcolor{LBGreen}T1218.003 & CMSTP \\
 \cellcolor{LBBlue}T1480 & Execution Guardrails \\
 \cellcolor{LBGreen}T1480.002 & System Checks \\
 \cellcolor{LBBlue}T1484 & Domain Policy Modification \\
 \cellcolor{LBBlue}T1484.001 & Group Policy Modification \\
 \cellcolor{LBBlue}T1486 & Data Encrypted for Impact \\
 \cellcolor{LBBlue}T1489 & Service Stop \\
 \cellcolor{LBBlue}T1490 & Inhibit System Recovery \\
 \cellcolor{LBGreen}T1543 & Create/Modify System Process \\
 \cellcolor{LBGreen}T1543.003 & Windows Service \\
 \cellcolor{LBBlue}T1547 & Boot/Logon Autostart Execution \\
 \cellcolor{LBRed}T1547.001 & Registry Run Keys/Startup Folder \\
 \cellcolor{LBGreen}T1547.004 & Winlogon Helper DLL \\
 \cellcolor{LBBlue}T1548 & Abuse Elevation Control Mechanism \\
 \cellcolor{LBBlue}T1548.002 & Bypass User Account Control \\
 \cellcolor{LBBlue}T1562 & Impair Defenses \\
 \cellcolor{LBBlue}T1562.001 & Disable/Modify Tools \\
 \cellcolor{LBGreen}T1562.009 & Safe Mode Boot \\
 \cellcolor{LBRed}T1564 & Hide Artifacts \\
 \cellcolor{LBRed}T1564.003 & Hidden Window \\
 \cellcolor{LBGreen}T1569 & System Services \\
 \cellcolor{LBGreen}T1569.002 & Service Execution \\
 \cellcolor{LBGreen}T1573 & Encrypted Channel \\
 \cellcolor{LBGreen}T1573.001 & Symmetric Cryptography \\
 \cellcolor{LBBlue}T1614 & System Location Discovery \\
 \cellcolor{LBBlue}T1614.001 & System Language Discovery \\
 \cellcolor{LBGreen}T1622 & Debugger Evasion \\
\bottomrule
\end{tabular}
\end{subtable}
\end{table}

\section{The LockBit group leak}
\label{sec:leak}
One of the most important aspects of the LockBit leak is that it provides many details about different aspects of the group and its operations. The leaked database contains the following 14 tables with data:
\begin{enumerate}
    \item \textbf{btc\_addresses:} A list of Bitcoin addresses used by the group's administration to siphon their payments.          
\item \textbf{builds:} Information about the victim, the ransomware that was deployed, and status
\item \textbf{builds\_configurations:} Details about the builds and the configurations used to activate specific modules of the ransomware.
\item \textbf{chats:} Negotiation chats with the victims.
\item \textbf{clients:} Victim information and links with builds.
\item \textbf{files:} metadata about the files that were exchanged on the platform
\item \textbf{invites:} Records regarding the invites to join the \$777 low-tier affiliate program of LockBit, a scouting side project of the group to recruit new affiliates. 
\item \textbf{migrations:} A summary of changes to the backend.
\item \textbf{news:} Generic news of the group
\item \textbf{pkeys:} The public keys used for encryption of the ransomware.
\item \textbf{socket\_messages:} A log of messages received by the platform, most likely used for attack detection.
\item \textbf{system\_invalid\_requests:} A log of invalid requests to monitor possible attacks to the platform.
\item \textbf{users:} a table with details about the affiliates and administrator of the platform, including credentials and contact information (TOX),
\item \textbf{visits:} A log of when each affiliate logged into the platform.
\end{enumerate}

\begin{figure*}[th!]
\centering
\includegraphics[width=\textwidth]{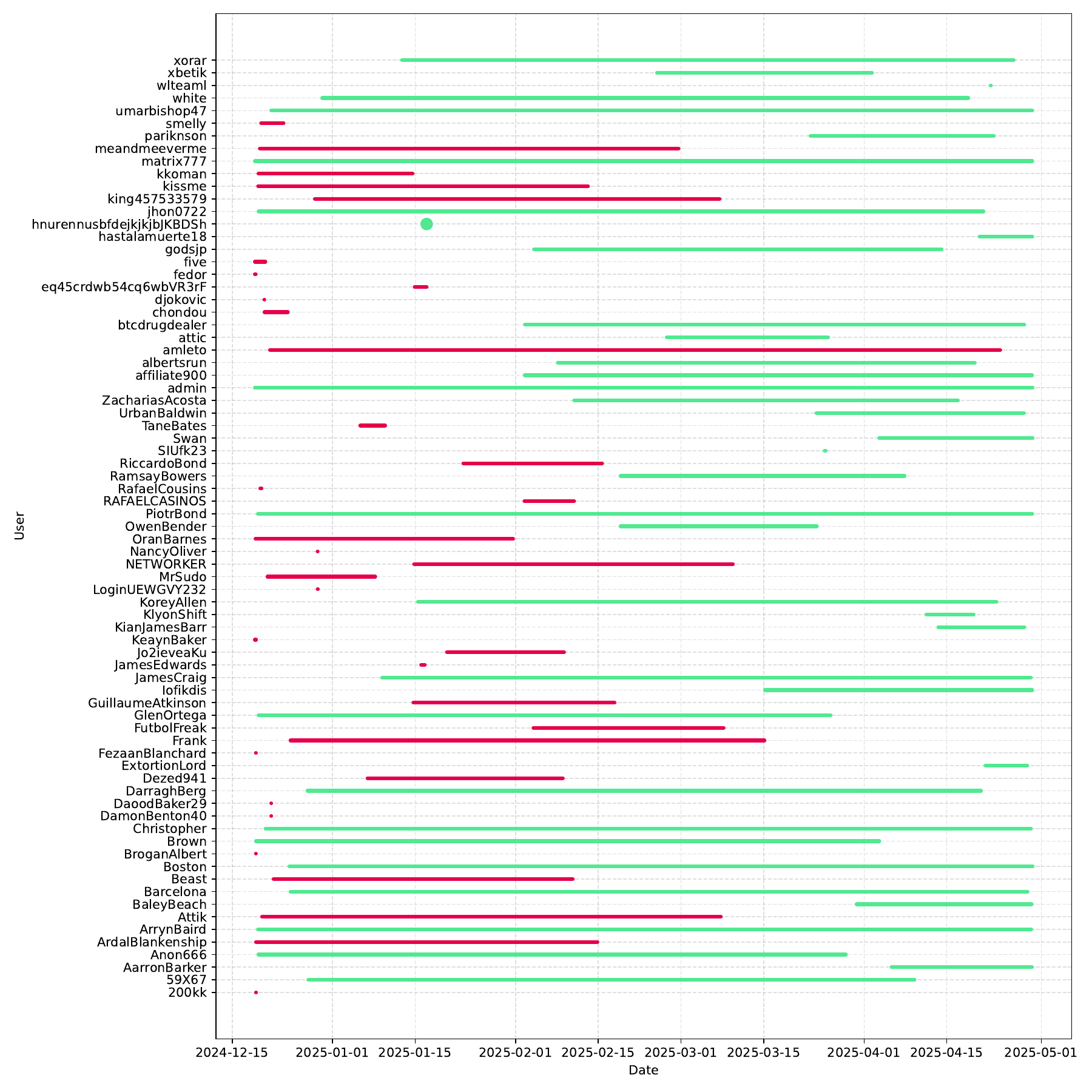}
\caption{Platform usage activity by LockBit members. Notation: Red users have been paused, green users were active when the database was leaked.}
\label{fig:logins}
\end{figure*}

Initially, the leaked database contains the list of users with their passwords in plaintext form, but also their TOX ID, where available. The table users contains 75 users, which are labeled as Verified (n = 5), Scammer (n=1), pentester (n=4), and newbie (n=62). Only two users, the admin and matrix777, do not have a tag, and one (king457533579) is labeled as 'ru target', which will be discussed in Section \ref{sec:incidental}. It should be noted that one user is labeled as 'pentester ?', but for the sake of clarity, we have included them along with the rest of the pentesters. Since the database allows us to extract their platform usage activity, which is illustrated in Figure \ref{fig:logins}.

Setting aside the fact that the passwords were stored in plaintext form on the platform, many of the members did not use secure passwords. For instance, there are several easy-to-guess passwords, e.g., Lockbit123, signifying that the group as a whole did not follow the best security practices. Finally, based on the username similarity with previously leaked usernames of affiliates \cite{nca}, we can assume that at least seven members of the previous platform, before Operation Chronos, continued in the revived platform.

\section{Behavioral Analysis of Attacker-Victim Interactions}
\label{sec:attackerbehavior}
The primary objective of this analysis is to identify the patterns and strategies used by the attackers during their interactions with victims. This interaction guide is commonly referred to as a playbook. In this specific scenario, we aim to analyze the leaked LockBit chat to identify recurring behaviors and linguistic structures by examining attacker messages within recorded conversations. 

\begin{figure*}
    \centering
\includegraphics[width=.9\textwidth]{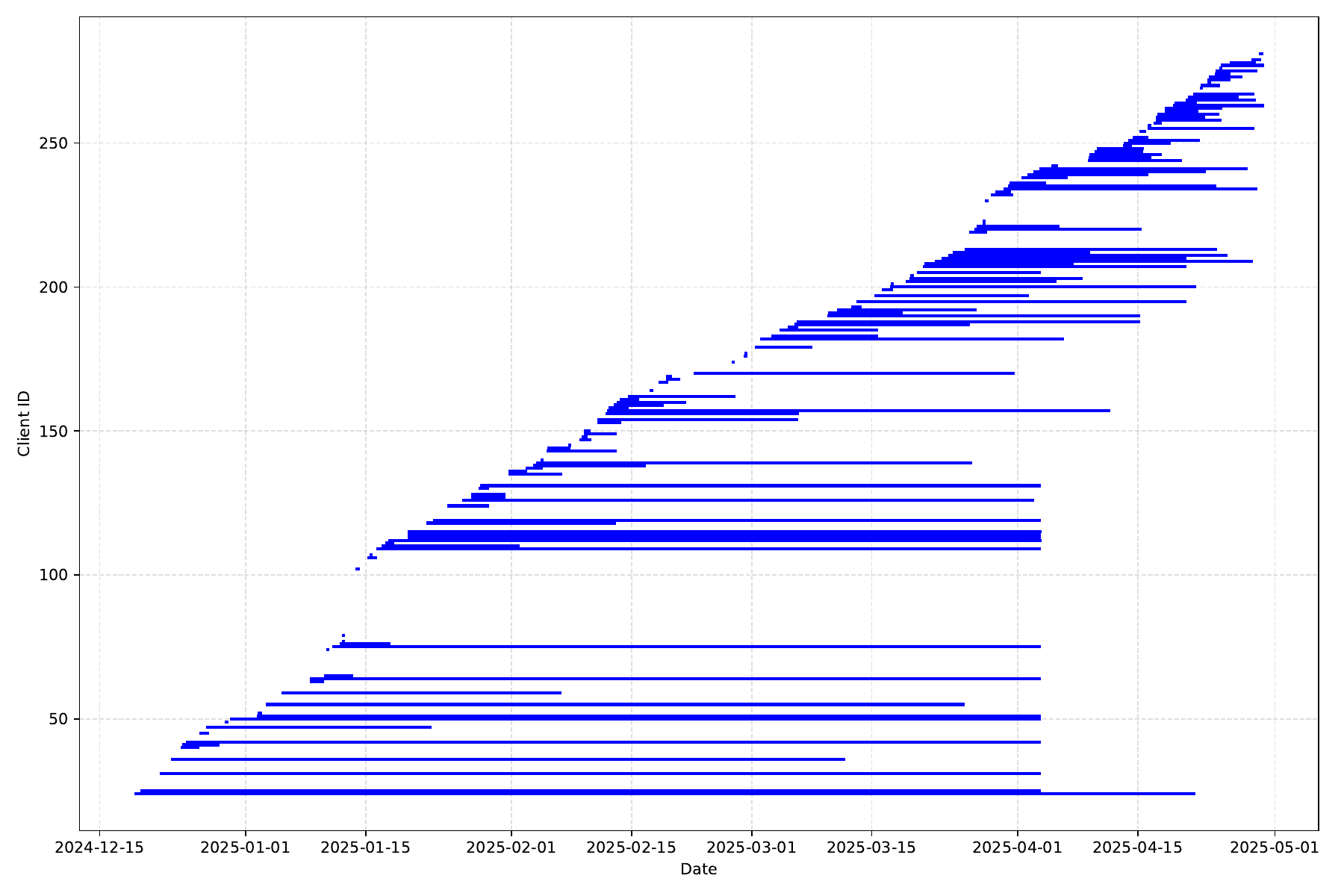}
\caption{Negotiation chat visibility duration.}
\label{fig:duration}
\end{figure*}

The database contains 208 negotiation chats, many of which do not contain victim interaction.
It is worth noting that the database logs the activity in the chat, even if it is simply checking whether the victim visited the chat. Based on that, we created Figure \ref{fig:duration}, which illustrates the victims' visiting patterns. The figure shows that many of the chats were visited for prolonged periods, hinting that many of the chats were exposed to researchers who repeatedly logged in to check for new information.     

\begin{figure}[th]
    \centering
    \resizebox{\linewidth}{!}{ 
    \input{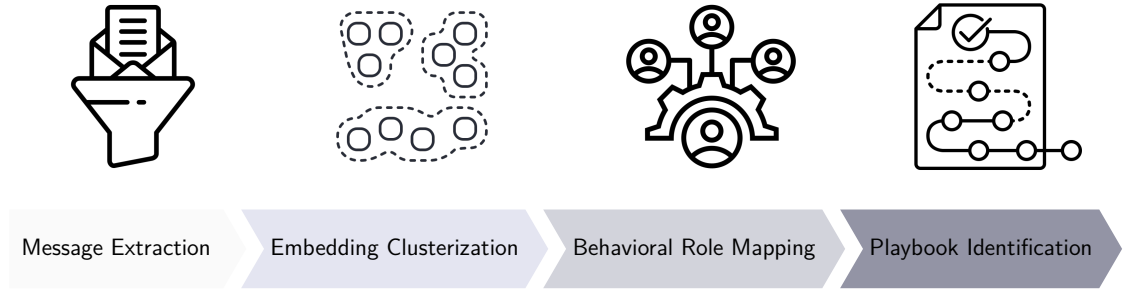}
    }
    \caption{Interaction diagram of Backend and Frontend functionalities.}
     \label{fig:graphic_summary}
\end{figure}

A pipeline is implemented using several natural language processing and machine learning techniques to analyze the behavioral strategies used by attackers. The process begins with the extraction of the messages, followed by semantic encoding using transformer-based embeddings to capture the meaning of individual messages and their clustering using the K-Means algorithm. Later, we utilize a large language model (LLM) to interpret each cluster and assign behavioral labels in a step known as behavioral role mapping. This task enables the identification of recurring communication roles. Finally, by examining the temporal sequence of labeled messages within segmented conversations, a behavioral graph is constructed to model typical interaction flows, which are the basis for the final behavioral playbook. The summary flow can be seen in Figure \ref{fig:graphic_summary}.

\subsection{Text Filtering and Preprocessing}

In what follows, for message extraction from all available conversations, only the messages sent by the attackers are selected for this analysis. This filtering ensures that the interactions of the victim do not dilute the behavioral patterns. This approach enables a clearer understanding of the strategies of the attackers, linguistic structures, and negotiation tactics. Identifying these repetitive behaviors is a crucial step toward developing automated detection methods and enhancing threat intelligence in ransomware negotiation scenarios. 

In the message extraction step, we preprocess the samples, aiming to standardize the language and reduce lexical variability. This task included text cleaning, converting to lowercase, removing irrelevant elements such as URLs and user mentions, and eliminating common stopwords. Additionally, lemmatization is employed to unify different grammatical forms of words. As a result, messages become easier to process and cluster, facilitating the identification of recurring patterns. An illustrative example of this preprocessing outcome is presented in Table \ref{tab:preprocessed_samples}.

\begin{table}[ht]
    \centering
    \rowcolors{2}{}{gray!10}
    \caption{Comparison between original and preprocessed messages}
    \label{tab:preprocessed_samples}
    \renewcommand{\arraystretch}{1.15}
    \begin{tabular}{p{8cm}p{6cm}}
        \toprule
    \textbf{Original Message} & \textbf{Preprocessed Message} \\
    \midrule
Yes, the amount in Bitcoin indicated above is final. The sooner you close the deal, the better. &
yes amount bitcoin indicate final soon close deal well \\
http://lockbitfskq2fxclyfrop5yizyxpzu65w7
pphsgthawcyb4gd27x62id.onion/r/vEtRey
ad1v\#QPZkIQsLLKABpUZckTe5MHuXuZU
i3GVBhXt6m8atjwo= Here download link &
download link \\
You can attach a few files for test decryption by packing them into an archive with zip, rar, tar, 7zip, 7z, tar.gz extensions of no more than 10 megabytes using the attach button directly in the chat.

If your archive weighs more than 10 megabytes, please use our file sharing service.
http://lockbitfss2w7co3ij6awox4xcuxx.onion
http://lockbitfsvf75glg226he5inkxx.onion
http://lockbitfskq2fxclyfropxx.onion
For security reasons we do not click on other links you send in chat.
Please wait for a reply, sometimes it takes several hours due to possible time zone differences.. &
attach file test decryption pack archive zip rar tar tar gz extension megabyte use attach button directly chat archive weigh megabyte please use file sharing service security reason click link send chat please wait reply sometimes take several hour due possible time zone difference \\
\bottomrule
    \end{tabular}
\end{table}

 \subsection{Embedding Clusterization}
The next step involved transforming the preprocessed data into vector representations to explore behavioral patterns. The goal of this step is to group similar messages and uncover recurring communication structures that may reflect underlying attacker strategies or shared objectives. To this end, we generated sentence-level embeddings using a transformer-based language model optimized to capture semantic similarity in short texts. Specifically, we used the all-MiniLM-L6-v2 model. This representation allows each message to be mapped into a high-dimensional vector where semantically similar sentences are positioned closer, which is valuable when identifying playbooks based on the free-form text input that attackers might use.

\begin{figure}[th]
    \centering
    \begin{subfigure}[t]{\columnwidth}
    \centering
\includegraphics[width=.8\linewidth]{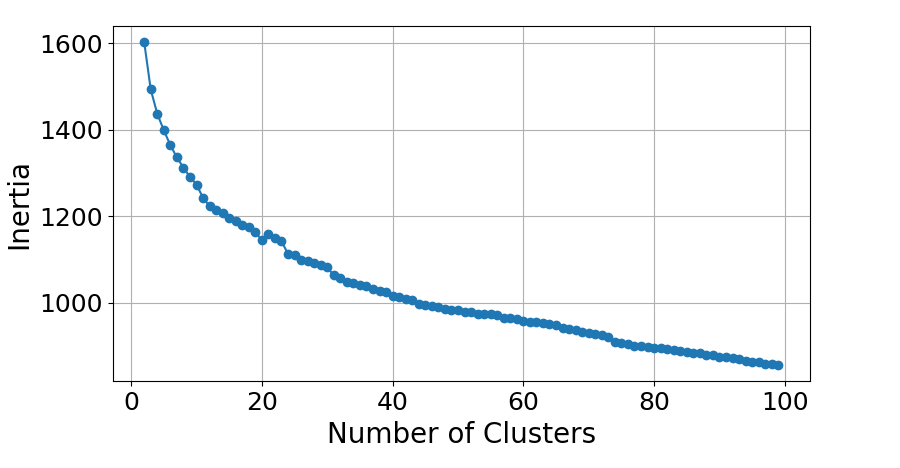}
    \caption{Elbow method.}
    \end{subfigure}
    \begin{subfigure}[t]{\columnwidth}\centering
\includegraphics[width=.8\linewidth]{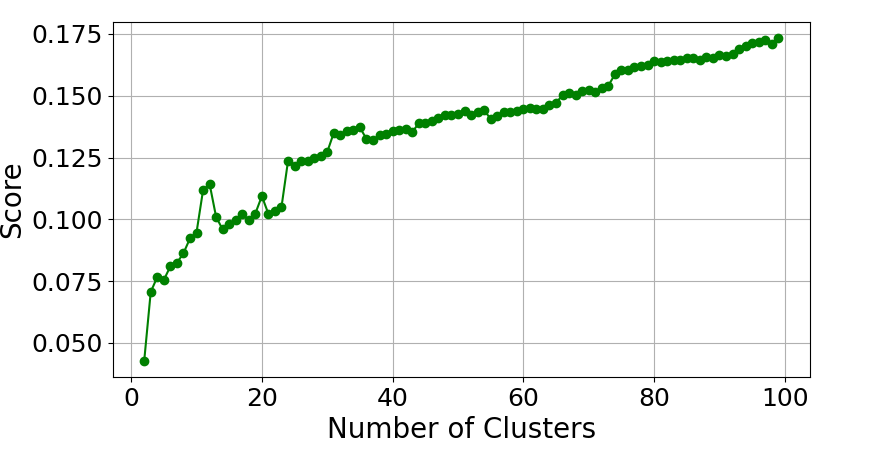}
    \caption{Silhouette score.}
    \end{subfigure}
    \caption{Detail of the outcomes of Elbow (top) and Silhouette (bottom) for the tested number of clusters.}
     \label{fig:elbow_silhouette}
\end{figure}

Making use of the semantic vector representations, we applied K-Means clustering to uncover latent structures in the data. To determine the optimal number of clusters, we evaluated configurations ranging from 2 to 100 using four standard metrics: inertia, silhouette score, Calinski-Harabasz index, and Davies-Bouldin index. The analysis indicated that approximately 24 clusters provided the best balance, with both the elbow method and the silhouette score peaking in that range; see Figure \ref{fig:elbow_silhouette}. In support of this, the 24-cluster configuration achieved the highest Calinski-Harabasz score and the lowest Davies-Bouldin score, suggesting it offered the most distinct and compact grouping for this specific dataset, as can be seen in Table \ref{tab:cluster_metrics}. 

\begin{table}[ht]
\centering
\rowcolors{2}{}{gray!10}
\caption{Comparison of clustering quality metrics for candidate cluster counts (22–26).}
\label{tab:cluster_metrics}
\scriptsize
\renewcommand{\arraystretch}{2.15}
\begin{tabular}{cccccc}
\toprule
\textbf{n\_clusters} & \textbf{Inertia} & \textbf{Silhouette} & \textbf{Calinski-Harabasz} & \textbf{Davies-Bouldin} \\
\midrule
22 & 1149.65 & 0.103 & 42.48 & 3.09 \\
23 & 1142.66 & 0.105 & 41.32 & 3.02 \\
\rowcolor{red!8} 24 & 1112.93 & 0.124 & 42.81 & 2.90 \\
25 & 1109.84 & 0.121 & 41.35 & 3.03 \\
26 & 1098.58 & 0.124 & 40.88 & 2.99 \\
\bottomrule
\end{tabular}
\end{table}

The next phase of the analysis focuses on interpreting the content of each cluster and mapping it to behavioral roles within attacker interactions. We employed an LLM, more precisely GPT-4 mini, to assist in generating concise and coherent summaries that capture the thematic essence of each group. The complete set of messages within each cluster was provided to the model to produce descriptive overviews of the underlying content.

We submitted the complete list of messages from each cluster to the LLM to assign behavioral labels using the prompt depicted in Figure \ref{fig:prompt}. Those requests were sent through the API, and the resulting labels are subsequently manually reviewed to ensure their accuracy and contextual relevance in the framework of attacker communications. Figure \ref{fig:bar_figure} presents the behavioral topics and corresponding cluster numbers for the ten most populated groups.

\begin{figure}[!th]
    \centering
        \footnotesize
        \begin{tabular}{|p{.95\columnwidth}|}   
            \hline 
            \cellcolor{red!8}
            \textbf{Prompt:}     Given the following list of short sentences, return one concise behavioral description for the group (3–6 words). Focus on the underlying intent or action (e.g., greeting, negotiating, threatening, sharing files). Return only the description\\
            \hline
        \end{tabular}
 \caption{Structure of the task prompt used in OpenAI's GPT.}
    \label{fig:prompt}
\end{figure}

\begin{figure}[!ht]
    \centering
    \begin{tikzpicture}
\begin{axis}[
    height=6cm,
    xbar,
    yticklabel style={font=\scriptsize},
    xticklabel style={font=\scriptsize},
    width=\columnwidth,
    xlabel={Number of Samples},
    symbolic y coords={
        Cluster 15,
        Cluster 6,
        Cluster 19,
        Cluster 22,
        Cluster 11,
        Cluster 7,
        Cluster 20,
        Cluster 0,
        Cluster 21,
        Cluster 4
    },
    ytick=data,
    nodes near coords,
    nodes near coords align={horizontal},
     every node near coord/.append style={
        font=\scriptsize,
        color=white,
        anchor=east,
        xshift=-4pt
    },
    enlarge y limits=0.1,
    xmin=0,
    xtick distance=20
]
\addplot [fill=blue, draw=none] coordinates {
    (80,Cluster 15)
    (86,Cluster 6)
    (92,Cluster 19)
    (96,Cluster 22)
    (103,Cluster 11)
    (118,Cluster 7)
    (120,Cluster 20)
    (120,Cluster 0)
    (125,Cluster 21)
    (152,Cluster 4)
};
\end{axis}
\end{tikzpicture}
\noindent
{\scriptsize
\textbf{Legend:} Cluster Number $\rightarrow$ Topic Label
\begin{itemize}[leftmargin=2em, itemsep=0.2em]
  \item Cluster 4: waiting for decryptor from tech team
  \item Cluster 21: negotiating prices
  \item Cluster 0: explaining solution time window
  \item Cluster 20: introduction and demands
  \item Cluster 7: decryption transaction
  \item Cluster 11: free decryptor test
  \item Cluster 22: negotiating prices
  \item Cluster 19: Threatening about losing data or sharing data
  \item Cluster 6: requesting payment
  \item Cluster 15: explaining commands for decryption
\end{itemize}
}

    \caption{Top 10 topics found in the analysed data.}
     \label{fig:bar_figure}
\end{figure}

As part of this manual verification, four clusters were excluded from further analysis because they exhibited mixed content, chat configurations, outliers (Russian messages), or agreeing messages, such as short messages with "okay", "ok", or "yes". These clusters are deemed either thematically insignificant or representative of isolated cases that did not reflect generalizable behavioral patterns. Nevertheless, the case of Russian messages will be discussed later. The remaining clusters are used to assess the relative distribution of behavioral patterns.

\subsection{Playbook Identification}

Using the set of clusters discussed above, we map individual messages to their corresponding behavioral categories and analyze them in terms of their temporal positioning within interaction sequences. This approach enables the reconstruction of a behavioral trajectory for each interaction, offering insights into the typical progression of attacker communication.

We conduct a temporal analysis to extract the behavioral playbook. Conversations are segmented based on the interaction of the attacker and the victim and the time at which the interaction occurred. Specifically, any message gap exceeding three hours was treated as the start of a new interaction, resulting in 148 distinct interactions identified.

\begin{figure}[!ht]
    \centering
\includegraphics[width=\columnwidth]{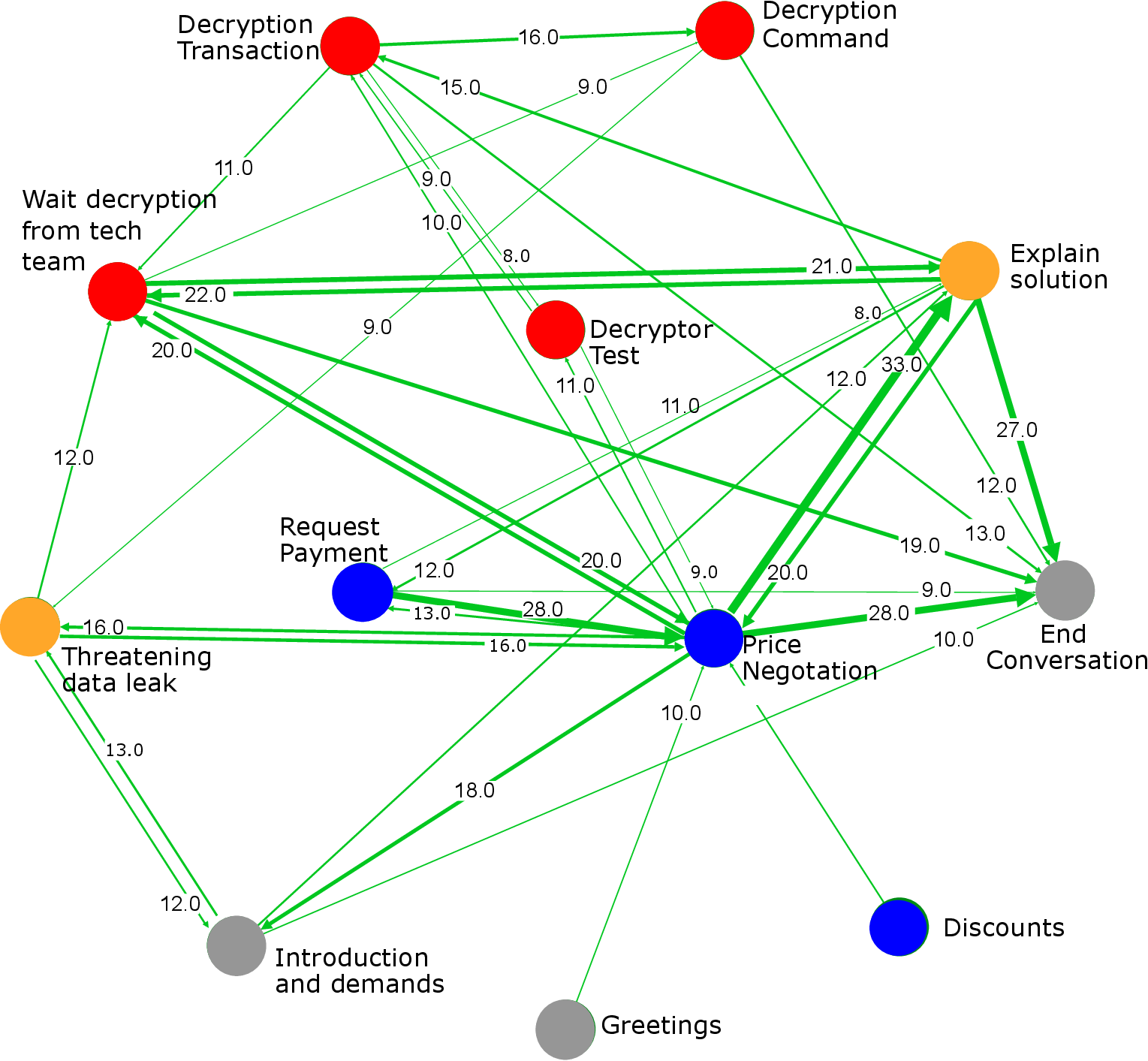}
    \caption{Behavioral flows observed in attackers. Notation, category related to: 
    \circlecolor{mygray}: Init/End Conversation,
    \circlecolor{red}: Decryption,
    \circlecolor{myorange}: General Information or Threatening, and
    \circlecolor{blue}: Ransom.
}
     \label{fig:intial_graph}
\end{figure}

Once the interactions are defined, each one is examined to determine the sequence of behavioral clusters it contains. Based on this information, we build a directed graph in which each node represents a behavioral theme, and each edge reflects the probability of transitioning from one theme to another. The weight of each edge represents the frequency of observed transitions between specific themes across the interactions. The resulting structure captures the behavioral flows observed in the interactions of attackers and serves as the foundation for the playbook. Figure \ref{fig:intial_graph} shows a raw diagram of transitions between themes. For clarity, we have removed nodes without edges.

Finally, to extract the behavioral playbook, we focused on identifying recurrent transition patterns across conversations. We established a top 15\% threshold for transitions with the highest probabilities to focus on the most frequent and significant connections within the graph. We selected this threshold empirically to balance coverage and interpretability: lower thresholds (e.g., top 5–10\%) risked omitting meaningful yet frequent transitions, while higher thresholds (e.g., above 20\%) introduced too much noise, making the resulting playbook less concise and actionable. 

\begin{figure}[!ht]
    \centering
\includegraphics[width=\columnwidth]{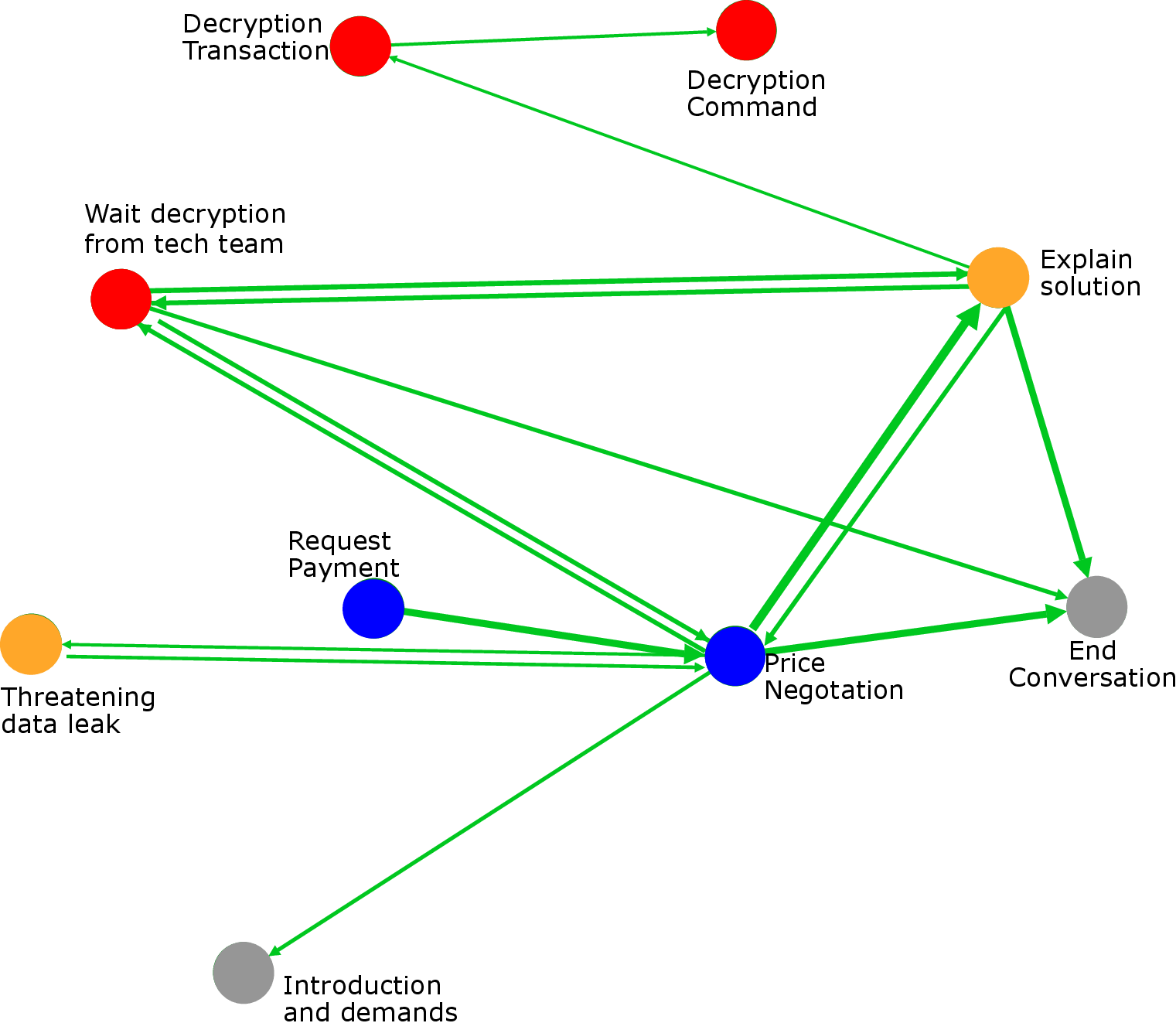}
    \caption{LockBit Playbook. Notation, category related to: 
    \circlecolor{mygray}: Init/End Conversation,
    \circlecolor{red}: Decryption,
    \circlecolor{myorange}: General Information or Threatening, and
    \circlecolor{blue}: Ransom.}
     \label{fig:playbook}
\end{figure}

The new graph represents the core behavioral dynamics by filtering out less frequent or incidental transitions, which are less likely to generalize across interactions. This approach ensures that the resulting playbook reflects the fundamental structure of the attacker dynamics, as illustrated in Figure \ref{fig:playbook}, highlighting the dominant behavioral pathways observed throughout the dataset. The analysis shows that the graph preserves nine distinct behavioral themes (nodes), indicating that low-value transitions were removed and a broad and meaningful range of key behavioral topics remains.  

It is essential to clarify the apparent discrepancy in the interaction patterns associated with the 'Introduction and Demands' node. While this node is represented as an entry point in the interaction sequence, it does not appear to transition to other nodes. This behavior is the result of two key considerations in the playbook reconstruction process. First, a considerable portion of conversations begin prior to the timestamp of the database dump used in our analysis. As a result, some of the earliest interactions are not captured in the dataset, leading to an underrepresentation of outgoing connections from this node. Second, our methodology applies a three-hour inactivity threshold to separate conversational interactions. This segmentation approach can fragment what would otherwise be continuous exchanges, particularly around the early stages of communication.

Additionally, the playbook is derived using a threshold to determine the significance of connections between nodes. Given the previously mentioned constraints, the number of transitions originating from the ``Introduction and Demands'' node may fall below this threshold, thereby limiting its representation in the final playbook visualization.

\subsection{Incidental findings}
\label{sec:incidental}
Going through the negotiations, there are several interesting findings. For instance, LockBit appears to be experimenting with shifting from Bitcoin payments to Monero. Indeed, some affiliates offered around 20\% discount should the payment be performed on Monero. We consider this shift to be rather important, as Monero transactions offer more privacy than Bitcoin \cite{li2019traceable}. However, it should be noted that the majority of operations still occur on the Bitcoin blockchain.

Moreover, despite the promises for full decryption of the victims' files, the chats showcase that this is not always the case. There are some chats in which the victims complain that their files could not be decrypted. LockBit affiliates claim that this was the result of other mechanisms blocking their encryption mid-process or the victims stopping the corresponding process. Due to the fact that in many instances, victim organizations hire experts for incident response and management, they may perform the negotiations. In one such event, the victim's negotiator appears to continue the fight with the affiliate, who eventually reveals that he has moved to LockBit after working in another group, disclosing the ID he had posted in a hacker forum. The exchange of affiliates is also manifested by the fact that in a negotiation chat, the victim is puzzled about whom he is talking to, as he was hacked by RansomHub, another ransomware group. However, the affiliate notifies the victim that practically the negotiations will be performed with LockBit, as the affiliate has moved, along with the victim's data.   

Interestingly, the negotiator from the victim's side asked for details to join the `business' or even to attack other organizations. Although one could consider this a form of delay, it was part of the negotiations that led to a payment, and, in fact, on some occasions, it occurred after the payment. It should be noted here that LockBit has, in the past, advertised the recruitment of affiliates, even in the form of providing valid credentials to access the VPN.

In terms of the requested ransom, it is apparent that the group examines the exfiltrated data to assess what the victim can pay. Indeed, the affiliates on several occasions refer their victims to their insurance contract, noting that it can cover their costs. Obviously, knowing the contract of their victims illustrates that LockBit specifically searched for it and based their requests on it. Likewise, on another occasion, the affiliate refers to the victim's bank statements and their balance, illustrating that they are well aware of the victim's financial capacity based on the exfiltrated data. 

Finally, it should be noted that there is a Russian victim organization and a Russian-speaking victim. The affiliates in both cases escalated the issue to the administration (the LockBit boss), who proposed providing free decryptors. In both cases, the administrator claimed that this issue was not the group's responsibility but rather the work of an infiltrator. In fact, in one case, the `boss' claimed that the key had been manually manipulated with a hex editor to prevent decryption. As a result, the victims claim that the decryptor does not work. 

\section{Cryptocurrency analysis}\label{sec:crypto}
Among all the information in the leaked database on LockBit's operations, the details related to cryptocurrency can be used to evaluate the economic impact of the group's activities and to extract their \emph{modus operandi} for collecting funds, moving them through their network, and performing money laundering operations. First, we provide an overview of the cryptocurrency information in Section \ref{sec:overview}. Next, we present the graph analysis in Section \ref{sec:graphanalysis}, while a summary of the outcomes is discussed in Section \ref{sec:results}.

\subsection{Cryptocurrency data}\label{sec:overview}
To analyze and evaluate the economic activities of the LockBit group, the entire Bitcoin blockchain up to May 31, 2025, was examined, encompassing more than 870,000 blocks and 1.12 billion transactions. Therefore, all outcomes and experiments presented in this study reflect the state of economic LockBit operations as of May 31, 2025, recognizing that subsequent transactions may alter the results.

The leaked database includes a table that contains nearly 60,000 Bitcoin addresses. However, only 19 of the 59,975 addresses ($\approx$0.03\%) have received funds. Moreover, by analyzing the conversations (chats) between the ransomware groups and their victims, 51 additional Bitcoin addresses can be identified. Specifically, since Bitcoin addresses have a specific pattern,  regular expressions were used to extract them, as also shown in \cite{wang2024automated}. These 51 detected addresses appeared in multiple conversations, resulting in only 25 unique addresses, as described in Table~\ref{tab:cryptocurrencyanalysis}. However, only 19 out of them ($\approx$76\%) have received funds. 

For the sake of clarity, from now on, the addresses included in the specific table of the leaked database are referred to as \emph{LBAs (LockBit Backend Addresses)}, while those extracted from the conversations are referred to as \emph{LCAs (LockBit Chat Addresses)}. Finally, the bitcoin addresses used to receive payments from the invitation program will be referred to as \emph{LockBit Bitcoin Invite Addresses (LBIA)}, to differentiate them from the Monero addresses that the program has.

\begin{table}[ht]
\centering
\caption{Overview of the cryptocurrency data available in the leaked database and chats.}
\label{tab:cryptocurrencyanalysis}
\rowcolors{2}{}{gray!10}
\begin{tabular}{lp{1.5cm}cc}
  \toprule
 & \textbf{Backend (from Database)} & \textbf{Chats} & \textbf{Invites}\\ \midrule
\textbf{\# Bitcoin Addresses} & 59,975 & 51 &2,338 \\ 
\textbf{\# Unique Bitcoin Addresses} & 59,975 & 25 & 2,338 \\ 
\textbf{\# Active Bitcoin Addresses} & 19 & 19 & 12\\ 
\textbf{Total BTC received} & 4.95974849 & 6.77328425 & 0.10036289\\
\textbf{Acronym used in this article} & LBA & LCA & LBIA\\ \bottomrule
\end{tabular}
\end{table}

\subsection{Graph-based approach}\label{sec:graphanalysis}
\begin{figure}[!htbp]
\centering
   \includegraphics[width=.8\columnwidth]{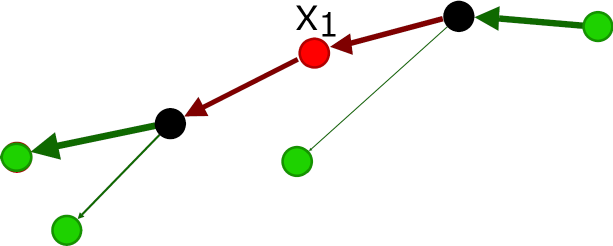}
    \caption{Example of 1-step address-transaction graph. Notation: 
    \circlecolor{red}: LockBit address,
    \circlecolor{mygreen}: Address, and
    \circlecolor{black}: Transaction.}
    \label{fig:add}
\end{figure}

To analyze the economic activities in which the LockBit cryptocurrency addresses have been involved, the address-transaction graph has been used, which shows the relationship between addresses and transactions in the corresponding blockchain (Bitcoin) \cite{jourdan2018characterizing}. Specifically, addresses and transactions are represented as vertices, while directed edges between addresses and transactions identify incoming relations (senders), and directed edges between transactions and addresses correspond to outgoing relations (receivers), as shown in Figure \ref{fig:add}. Moreover, both nodes and edges can be enriched with additional attributes, such as labels, amounts, fees, and timestamps. 

A similar graph-based approach has been successfully used to extract commonalities and patterns in previous works on mixer operations \cite{zola2025topological, moser2013inquiry}, to analyze spreading behaviors and similarities among ransomware families \cite{zola2024unveiling, c0c2adb7f0234e86a84d81251266913a}, and to evaluate the effectiveness of sanctions in the crypto ecosystem \cite{zola2024assessing}.

To build these graphs, a starting point needs to be defined (X$_1$), as well as the number of exploring steps $n$. This parameter specifies the number of transactions (both backwards and forward) to explore from a selected starting point. Therefore, the graph will include all the paths that originate from or lead to the starting point, with a maximum length of $2n$. In this article, LCAs are used as the starting points (X$_1$) for building the address-transaction graphs. This is because the chats clearly indicate that these addresses (LCA) are used by the group to request ransom from their victims, highlighting their central role in the investigations. Thus, the number of exploration steps of the graph ($n$) is set to two.

\subsection{Behavioral pattern results}\label{sec:results}
\begin{figure}[!ht]
\centering
   \begin{subfigure}[b]
   {\linewidth}
   \centering\includegraphics[width=.8\linewidth,trim={0 0 5cm 0},clip]{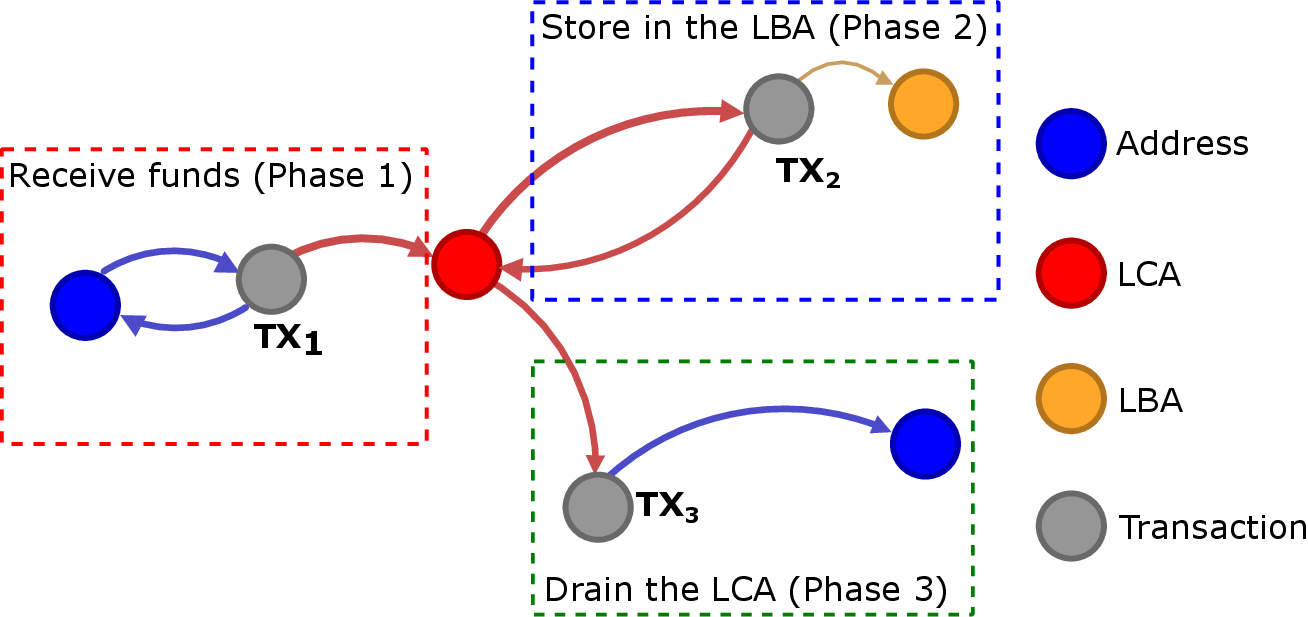}
    \caption{\textit{Pattern A:} three phases.}
    \label{fig:phases}
\vspace*{4mm}
\end{subfigure}
   \begin{subfigure}[b]{\linewidth}
   \centering\includegraphics[width=.8\linewidth,trim={0 1.5cm 4.7cm 0},clip]{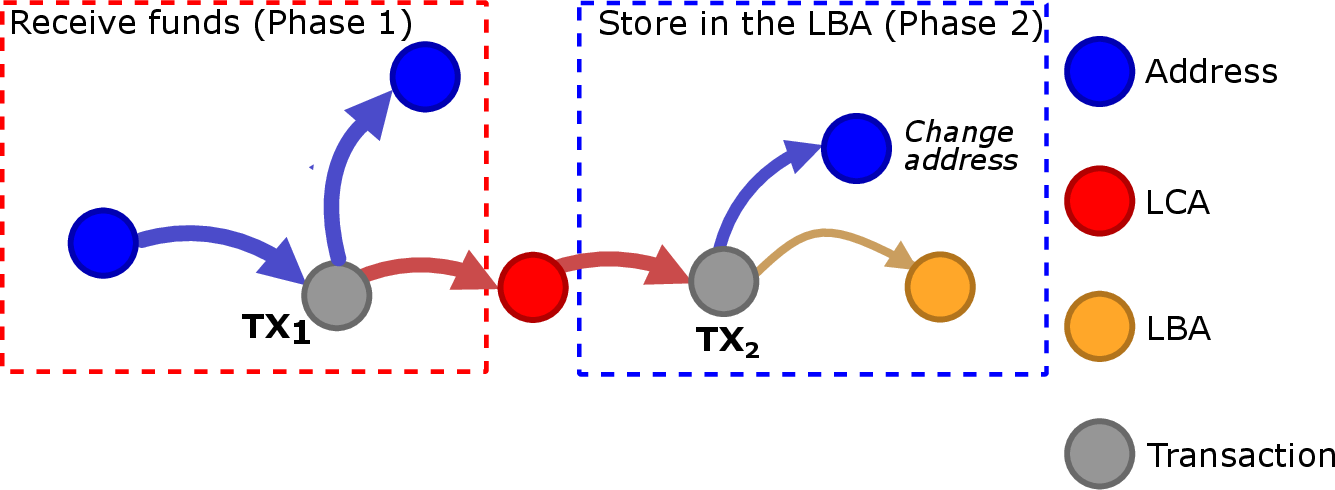}
    \caption{\textit{Pattern B}: two phases.}
    \label{fig:phases2}
\end{subfigure}
\caption{\textit{Modus operandi} extracted by analyzing address-transaction graphs of the LockBit addresses. Notation: 
    \circlecolor{mygray}: Transaction,
    \circlecolor{red}: LCA,
    \circlecolor{myorange}: LBA, and
    \circlecolor{blue}: Address.
    }
\label{fig:phases_all}
\end{figure}

The resulting address-transaction graphs have been carefully analyzed, and in 11 of the 19 available cases, they show three main phases, as illustrated in Figure~\ref{fig:phases}. In the first phase, obviously, the LCA receives funds, presumably from the ransomware victims ($TX_1$ in Figures \ref{fig:phases} and \ref{fig:phases2}). Then, in Phase 2, a small portion of the incoming is sent and stored in the LockBit Backend Addresses (LBA) through a single transaction ($TX_2$ in Figures \ref{fig:phases} and \ref{fig:phases2}); and a final phase in which the rest of the amount is sent in further transactions (can be more than one) until the LCA is completely drained ($TX_3$ in Figure \ref{fig:phases}). In 6 of the 19 available cases, the address-transaction graphs do not present Phase three, since in Phase two, a new address is used as the change address, generating a distinct flow, as the one depicted in Figure \ref{fig:phases2}. Finally, three cases do not follow any of the mentioned patterns.

\textbf{Phase 1: Receive funds.}
All 19 LCAs collectively received a total of 6.77328425 BTC. A closer analysis reveals that nine LCAs received funds from only one transaction, seven LCAs were funded through two transactions, and three LCAs received funds via three transactions each (the ones that do not follow any of the patterns indicated in Figure \ref{fig:phases_all}). Furthermore, in all these input transactions $TX_1$ (excluding 1), the sender was a single unique address. Yet, in six cases, this source address (sender) was associated with known cryptocurrency exchanges, highlighting potential reliance on centralized platforms to move funds into the LockBit ecosystem. This indicates a relatively low volume of transactional activity per address, which may suggest deliberate efforts to obfuscate or compartmentalize fund flows, as well as the victim's strategy to create a new address to send the ransom.

\textbf{Phase 2 - Store in the LBAs.}
In this phase, a small portion of funds is sent from each LCA to a specific LBA in a single transaction. This occurred in 16 out of the 19 cases, and in all of these cases, distinct LBAs are used. It should be noted that the remaining three LBAs also received funds from a unique sender; however, it has not been identified as LCA. This approach converts the LBAs into mainly recipient addresses, meaning they receive funds without sending them in turn.

A deeper analysis of the transaction flows reveals that LBAs are involved mainly in two transactional patterns, as illustrated in Phase 2 of Figure \ref{fig:phases_all}. 
More precisely, Phase two of pattern A shows that the LCA still sends funds to an LBA, but the change is returned directly to the LCA sender (Figure \ref{fig:phases}), opening the way to Phase three. On the other hand, Phase two of pattern B shows that the LCA sends funds to an LBA and another unlabeled address (Figure \ref{fig:phases2}), also known as a change address or LBA sibling. The pattern A occurs in eleven out of 16 cases, while the pattern B occurs in the remaining six cases.

Before concluding the analysis of Phase two, the change addresses identified in Pattern B were also examined. Specifically, we constructed the address-transaction graph in three steps ($n$ equals 3), using the change addresses as starting points. This approach did not reveal any consistent patterns. 
However, we found that in some cases, LCAs sent a share of the ransom immediately to an exchange, exponentially increasing the complexity of tracing the money flow. More precisely, three exchanges were used, with a total of approximately 1.784 Bitcoins in ransom being cashed out immediately, representing more than a quarter (26.34\%) of the total funds received by these addresses (Table \ref{tab:cryptocurrencyanalysis}).

Finally, by analyzing the amounts involved in all the Phase two transactions of both Patterns, a clear \emph{modus operandi} emerges. In fact, although the amount sent by the LCA varies in all cases, the distribution of the output is quite stable. Specifically, the LBA always receives an amount between 19\% and 20\% of the input value, while the remaining 80\% to 81\% is sent to a change address or returned to the sender. This allowed the LBAs that are involved in Phase two (16 of the 19) to save 0,66671708 BTC.

\textbf{Phase 3: Drain the LCA (only in Pattern A).}
Among the 11 cases that follow the pattern A, only two leave small amounts of funds in the LCA, while the remaining nine completely drain the address. Specifically, in six cases, the drain is performed through a single transaction to a unique output address. In three cases, two transactions are used for draining, each directed to a unique output address. In the remaining two cases, three and five transactions are performed to drain the LCA, respectively.

\begin{figure*}[!htbp]
\centering
   \includegraphics[width=0.85\linewidth]{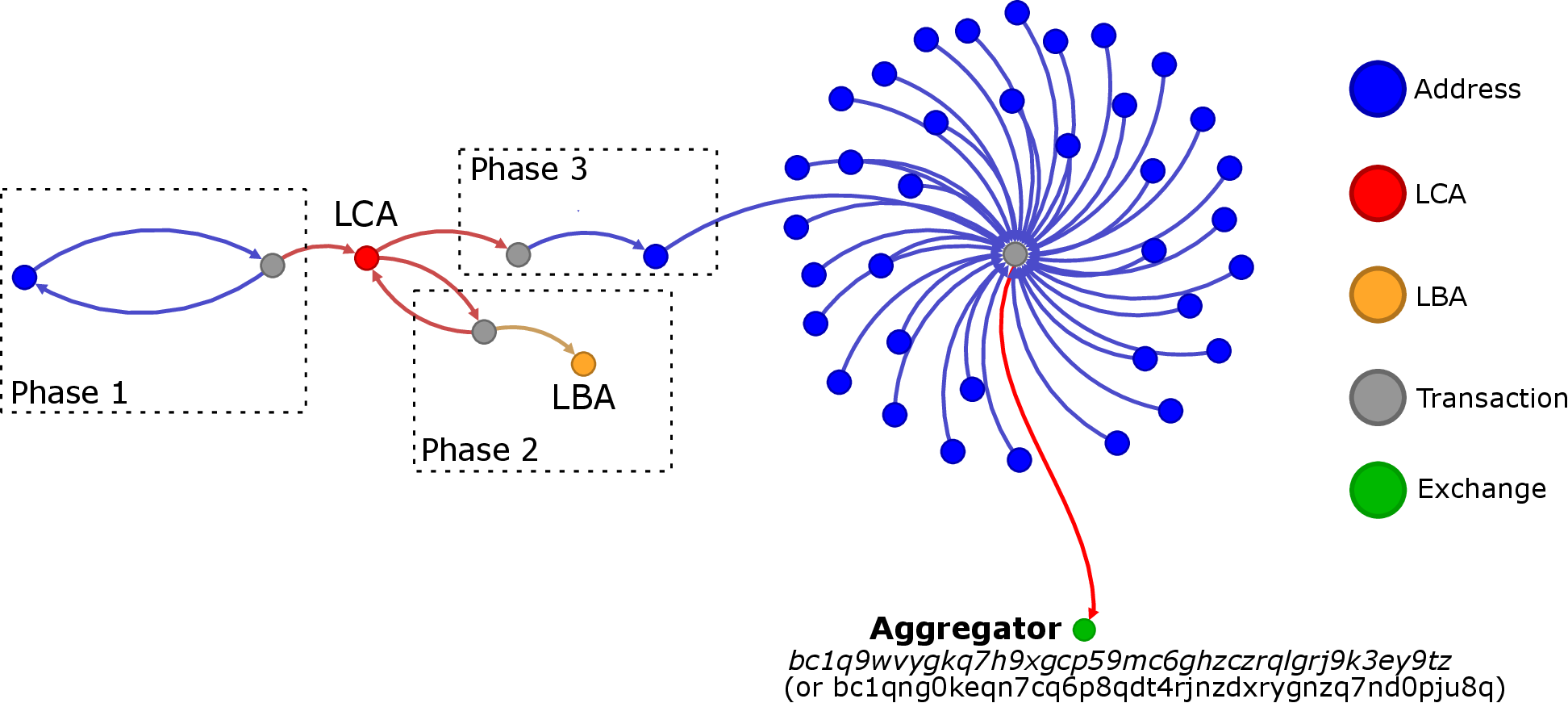}
    \caption{Aggregation network discovered by analyzing addresses involved in Phase three.}
    \label{fig:drain}
\end{figure*}

A further analysis of the draining activities, conducted by building address-transaction graphs using the addresses discovered in Phase three as starting points, reveals that in seven out of eleven cases, the Phase three addresses are subsequently involved in a single transaction in which their funds are combined with those from many other addresses and aggregated into a single destination, as illustrated in Figure \ref{fig:drain}. In particular, in all these cases, two specific addresses act as aggregators: \textit{bc1q9wvygkq7h9xgcp59mc6ghzczrqlgrj9k3ey9tz} and \textit{bc1qng0keqn7cq6p8qdt4rjnzdxrygnzq7nd0pju8q}. Upon analyzing the entities associated with these two aggregator addresses, we discovered that they are controlled by two concrete known exchanges. The first address was first used in October 2023 and remains active, continuing to participate in transactions to receive and send funds. Overall, it participated in almost 80k transactions, receiving an overall amount of $\approx$213,745 BTC (\$22,283M) and sending $\approx$213,469 BTC, keeping an active balance of 265 BTC. Similar numbers are shown by the second address that was used for the first time in June 2021, and it is involved in more than 76,000 transactions, receiving an overall amount of $\approx$283,674 BTC and sending $\approx$283,486 BTC (active balance of 183 BTC). This ongoing activity highlights the central role of exchanges in operations related to the LockBit ransomware.

\begin{figure}[!htbp]
\centering
   \includegraphics[width=.8\columnwidth]{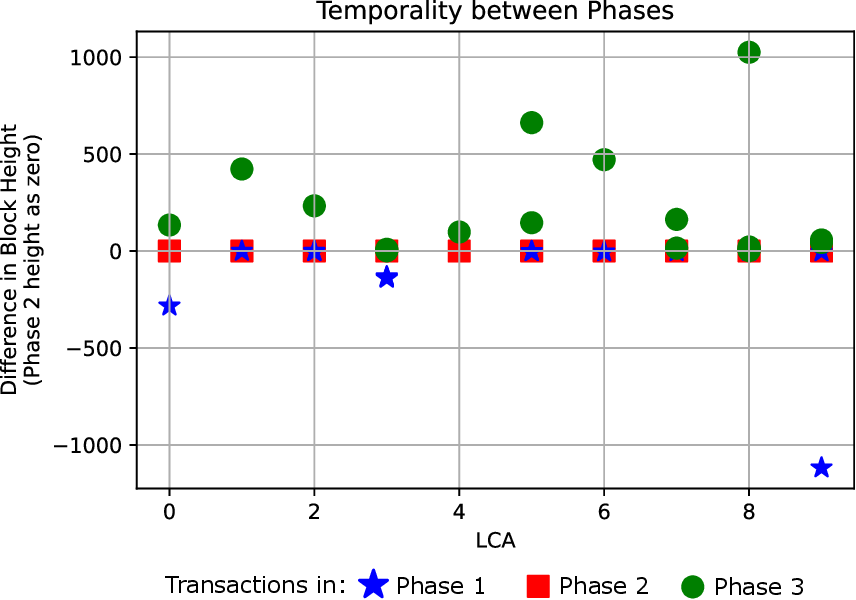}
    \caption{Height difference between the transactions in the different phases, considering transactions in Phase two as baseline (0).}
    \label{fig:temporality}
\end{figure}

Figure~\ref{fig:temporality} shows the block height differences (temporality) between all transactions involved in all the Phases of Pattern A. In particular, transactions belonging to Phase two are considered as the baseline (0 on the axes). This allows us to observe the time that passes between ransom collection (Phase 1), saving the funds to the LBA (Phase 2), and finally draining the LCA (Phase 3) in preparation for the money laundering operation. The figure shows that in seven cases, the transactions in Phase 1 and Phase 2 occur consecutively, with only a minimal delay. This supports the hypothesis that, once the funds are received from victims, the group aims to secure a portion of them in the LBA as quickly as possible. On the other hand, transactions in Phase 3 are performed immediately after those in Phase 2, in just four cases. In the other cases, they are performed with major delays, reaching also 1,000 block height difference ($\approx 7$ days). This supports the hypothesis that, before completely draining the LCA, the group wants to ensure that no other victims intend to send money to this LCA, so that the address can be ``retired'' and no longer be used in future operations.

Regarding the LBIA, there are only 12 active addresses, with most of them (8) having only one transaction, the received payment of \$777 for the registration to the affiliate program. The remaining four addresses were drained, leaving a zero balance, as all their funds were sent to the same address.

\section{Lessons Learned and Conclusions}
\label{sec:conclusions}
The unprecedented disclosure of LockBit's internal database offers a rare longitudinal perspective on a top-tier ransomware-as-a-service operation \cite{casino2025not}. Over the course of five years, the group has continuously refined its tooling, shifting from the comparatively noisy ``ABCD'' precursor to the stealth-centred LockBit 4.0 lineage. Each iteration demonstrates a deliberate response to defenders' counter-measures: harder-to-detect loaders, faster lateral-movement modules, and a data-theft stage that now rivals the encryption routine in strategic importance. The timeline reconstructed in this article underscores an uncomfortable but straightforward lesson: adversaries iterate as quickly as the security community reacts, and the lead time between a novel capability's introduction and its adoption by affiliates is measured in weeks, not months.

Nevertheless, the defenders are not the only ones making mistakes. Even ransomware groups make grave security mistakes that can expose them. Our behavioral study of the leaked negotiation chats provides an attacker-centric view of their \textit{modus operandi}. By clustering more than one hundred conversation threads, we understand that LockBit maintains a de facto playbook that can be disseminated to affiliates as readily as its binaries. Conversational monitoring, looking for the typical transition from automated requests to explicit threats, could therefore become an early-warning control point in extortion response workflows.

The cryptocurrency-flow analysis reveals an equally disciplined financial back-end. LockBit Chat Addresses act as transient buffers that split, almost immediately, incoming ransom payments in a near-constant 20/80 ratio, channeling the first part to long-lived storage addresses while forwarding the remainder through a mesh of change addresses before aggregation. The smaller proportion is likely to be the share taken by the group as a profit and to finance further operations. In fact, this 20/80 ratio matches the promises of the group to its affiliates in its new release \cite{lockbit_new}. On the other hand, the highest amounts are aggregated in two high-volume collector addresses before likely being sent to the affiliate. Specifically, these two collectors appear to belong to two distinct exchanges, once again highlighting their central role in ransomware operations and making it more difficult to trace the funds back.

Taking into account the findings as a whole, LockBit acts as a vertically integrated criminal enterprise whose resilience derives from tight technical iteration, a consensus social engineering methodology, and a cash-out infrastructure engineered for scale  \cite{patsakis2024malware}. Defenders must therefore match that integration with equally multidisciplinary counter-strategies, combining rapid patch hygiene, behavioral analytics, and coordinated financial intelligence, among others, to avoid being victims of such operations  \cite{mcintosh2023applying,stop_lockbit,casino2025unveiling}. Nevertheless, it is apparent that victims can still be exposed, even if they have paid the ransom, e.g., having their negotiations exposed, discussing how their ransom payments would be made to bypass regulatory audits, which can significantly backfire for them. Additionally, the chats reveal that even if the victims pay, there are several cases where they are unable to recover their data. 

Finally, the handling of Russian victims shows that the group does not want to interfere with them. This justifies the provision of free decryptors and the excuses given, e.g., infiltration from foreign agencies. After all, the group leader is alleged to be Russian and is being prosecuted by the USA~\cite{justice}.

Future work will extend the temporal horizon of both the communication and blockchain datasets to extract more intelligence and analyze potential attribution and collaboration among other well-known ransomware, fostering the elaboration of effective defense strategies. At the same time, visual analytics techniques can be integrated to assess the findings using defined metrics, extending beyond basic visualizations. By leveraging these methods, we aim to enhance investigative processes, generating an intelligence framework able to quickly detect and extract comprehensive information about groups' \textit{modus operandi}.  

\section*{Acknowledgment}
This work was partially supported by the European Commission under the Horizon Europe Programme, as part of the project SAFEHORIZON (Grant Agreement No. 101168562). This work was partially supported by Ministerio de Ciencia, Innovación y Universidades, Gobierno de España (Agencia Estatal de Investigación, Fondo Europeo de Desarrollo Regional -FEDER-, European Union) under the research grant PID2021-127409OB-C33 CONDOR, and by AGAUR with the project ASCLEPIUS (2021SGR-00111). Fran Casino was supported by the Spanish Ministry of Science and Innovation under the ``Ramón y Cajal'' programme (RYC2023-044857-I).

The content of this article does not reflect the official opinion of the European Union. Responsibility for the information and views expressed therein lies entirely with the authors.

\end{document}